# A Framework For Gait-Based User Demography Estimation Using Inertial Sensors

Chinmay Prakash Swami
Department of Mechanical and
Aerospace Engineering
University at Buffalo
Buffalo, US
chinmayp@buffalo.edu

*Abstract*—Human gait has been shown to provide crucial motion cues for various applications. Recognizing patterns in human gait has been widely adopted in various application areas such as security, virtual reality gaming, medical rehabilitation, and ailment identification. Furthermore, wearable inertial sensors have been widely used for not only recording gait but also to predict users' demography. Machine Learning techniques such as deep learning, combined with inertial sensor signals, have shown promising results in recognizing patterns in human gait and estimate users' demography. However, the black-box nature of such deep learning models hinders the researchers from uncovering the reasons behind the model's predictions. Therefore, we propose leveraging deep learning and Layer-Wise Relevance Propagation (LRP) to identify the important variables that play a vital role in identifying the users' demography such as age and gender. To assess the efficacy of this approach we train a deep neural network model on a large sensor-based gait dataset consisting of 745 subjects to identify users' age and gender. Using LRP we identify the variables relevant for characterizing the gait patterns. Thus, we enable interpretation of non-linear ML models which are experts in identifying the users' demography based on inertial signals. We believe this approach can not only provide clinicians information about the gait parameters relevant to age and gender but also can be expanded to analyze and diagnose gait disorders.

## I. INTRODUCTION

Among various biometric measures such as fingerprint, facial features, and iris, gait is emerging as a measure for identifying individuals and their characteristics. Various clinical and biomechanical studies have shown the uniqueness in the gait of each individual due to the interplay between various muscles and joints. Furthermore, it has been shown that gait is fairly consistent and is not easily changed for a healthy person which makes it a robust measure to differentiate one individual from others[1].

Gait characteristics have been analyzed in clinical settings through visual analysis or self-administered questionnaires. Whereas in research environments, motion capture systems have been used to evaluate the effects of age and gender on gait characteristics. Even though these methods are efficacious in evaluating gait, they are impractical for routine clinical use as these methods require skilled technicians and often involve a time-consuming setup process [2]. Gait is majorly characterized by using wearable systems such as Inertial sensors or via vision-based techniques such as images or video recordings [1,3]. Although, video camera-based gait recognition systems have shown promising results, often more than one camera is used to overcome the limitation of occlusion [1,3].

Wearable sensor-based gait recognition has emerged as an alternative to vison-based systems. These sensors, also known as Inertial Measurement Unit sensors (IMU), often comprise accelerometers to measure the movement of individuals' body segments whereas gyroscopes are used to measure angular velocities of these segments [4]. IMU sensors are effective in human gait analysis for various clinical applications to monitor and diagnose diseases [5]. However, accurate information about users' demography such as age and gender is vital for improving and maintaining the effectiveness of healthcare systems. Features extracted from inertial signals have been shown to contain discriminative features that aid in the estimation of age and gender [3]. Moreover, various studies have established variation in the gait of individuals belonging to different gender and age group. Therefore, since gait patterns vary with changes in age and gender, wearable sensors can be leveraged to effectively identify age and gender [2, 3, 5].

Riaz *et al.* were the first to establish the effectiveness of inertial sensors in recording gait characteristics for predicting age. Using the random forest algorithm, Riaz *et al.* achieved high classification accuracy [6]. However, research on gait-based age and gender estimation are still in its infancy. Furthermore, the use of wearable sensors for characterizing gait to estimate age and gender has not been widely explored [5].

Deep learning-based approaches have been popular in the clinical biomechanics area and have been successfully used to provide insights into the nature of human gait. Artificial Neural Networks and Support Vector Machines have been successfully used to establish uniqueness in the gait patterns, identify fatigue and successfully establish the effect of age and gender on gait. In clinical gait analysis, several approaches using machine learning have been proposed to categorize gait patterns and aid clinicians in identifying Parkinson's disease, cerebral palsy, multiple sclerosis, and brain trauma. Furthermore, these methods have also been applied to identify pathological gait conditions such as bone fractures or ligament injuries [7]. However, although these methods are efficacious, most of these approaches suffer from the limitation of being a black-box approach. This leads to the absence of reasons justifying the models' decisions which leads to reduced clinical acceptance. To overcome this challenge, Layer-wise Relevance Propagation (LRP) has been proposed for interpreting the non-linear models developed to solve problems in neuroscience, bioinformatics, and physics. Horst *et al.* leveraged LRP to increase the transparency of the Deep Neural Network model used for the classification of gait

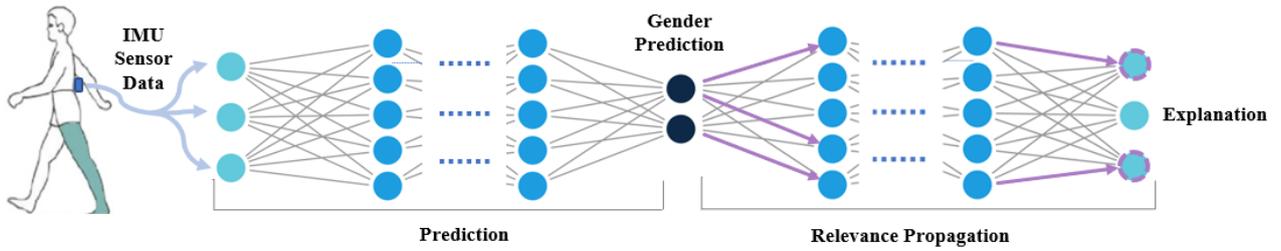

*Figure 1 Overview of proroposed framework for predicting user demography based on inertial measurements of gait.*

patterns by enabling the understandability and interpretability [7].

In this paper we propose a framework that leverages the non-linear nature of neural networks and the LRP technique to not only predict the user demography such as gender-based on the gait but also overcomes the black-box nature of these models, thus facilitating the identification of features highly correlated with gender. Such information can be utilized to design and develop effective gait-based user demography identification methods.

## II. RELATED WORK

To promote research on gait-based user demography estimation using wearable sensors, a challenge was introduced in a top-ranked conference (The 12th IAPR International Conference on Biometrics (ICB), Greece, 2019). The participants used two OU-ISIR datasets consisting of gait data from 745 subjects[8]. 18 teams participated in this challenge where 32 algorithms for gender estimation and 35 algorithms for age estimation were proposed[5].

### A. Proposed Feature Extraction Approaches

The submitted solutions employed signal pre-processing, feature extraction, and then estimated age or gender based on the processed signals. Fourier transformation of the accelerometer and gyroscope signals were used to train the ML classifiers. Statistical features have also been extracted from the signals to train the ML models. Some of the most commonly used statistical features are mean, standard deviation, variance, mean squared error, and various statistical entities applied to auto-correlation and auto-variance. Biomechanical features such as gait cycle; step length were also leveraged to improve the performance of the predictive models. Some solutions have also normalized the signals before using them for training the ML classifiers. Accelerometer data has also been used to compute PCA rotation matrix. Using the PCA rotation matrix, accelerometer and gyroscope signals were rotated which helped in removing the effects of gravity on accelerometer signals. Furthermore, the sliding window technique with some degree of overlapping has also been used for preprocessing the accelerometer and gyroscope signals. Lastly, some approaches employed no pre-processing steps and no features were extracted. The ML classifiers were trained on the RAW IMU signals[5].

### B. Proposed Classifiers

Both classical ML methods, as well as deep learning-based approaches, have been leveraged to estimate the age and gender based on the gait.

For estimating age based on gait patterns, various Deep Neural Networks (DNN) solutions have been proposed which varied based on the type of activation functions and optimizer used. ResNet has also been used for predicting age. A modified version of ResNet was proposed where Gated Recurrent Unit (GRU) was added between the last convolutional layer ResNet block and the dense layer. Bi-Directional Long Shot-Term Memory (BLSTM) has also been proposed with modified loss functions and different accumulation methods such as arithmetic mean and geometric mean. Apart from deep learning methods, ensemble-based methods such as random forest regression models have also been proposed. Classical machine learning models such as Support vector machines and decision regression trees have also been used to predict age based on the inertial signals. Lastly, the K-Nearest Neighbor method has also been employed to estimate the age based on the gait features.

Similar methods have also been used for estimating the gender based on the gait of an individual.

Although various deep learning-based solutions have been proposed developing gait-based estimators for age and gender, the focus has not been on the interpretability of the models. Furthermore, the solutions leveraging models which offer some degree of interpretability such as random forest or support vector machines were not exploited to deduce reasons behind conclusions. Such deductions play a vital role in not only troubleshooting and verifying the biomechanical correctness of the developed models but also help in designing more robust and novel gait-based estimators. Horst *et al.* leveraged LRP to decompose the predictions of an artificial neural network trained on joint angles of lower limbs. Therefore, the LRP technique led to the identification of relevant variables necessary for the characterization of the gait patterns of a certain individual. Due to the identification of relevant features, Horst *et al.* established the importance of symmetries and asymmetries between right and left body movements for the identification of individuals. As for differences between individuals, age and gender are an inherent feature of human gait, robustness and accuracy is essential which led the authors to also provide results indicating the higher performance of non-linear methods such as DNN's compared to linear methods[7].

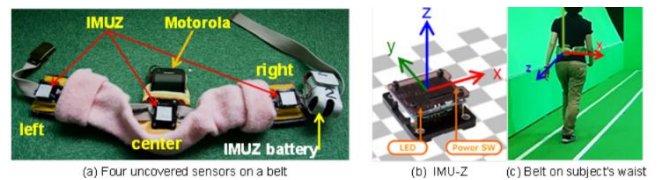

*Figure 2 IMU Sensor locations and axis orientations.*

## III. METHODS

After acquiring the consent for using the inertial dataset, the dataset was downloaded and the variation with 744 subjects and level-walk data was extracted [8]. This dataset comprised of accelerometer and gyroscope data recorded by the center sensor shown in Figure 2 which had a sampling frequency of 100Hz. Figure 1 illustrates the steps involved in the proposed framework.

*A. Data Preprocessing*

To eliminate noise from the gyroscope and accelerometer signals, a moving average filter with a window size of 10 was used. 1-second signals generated by the sensor for each axis were extracted and horizontally stacked to generate features. The data set was further divided into 3 parts; Training, Validation, and Test. The mean and standard deviation for each feature in the training dataset was computed and were used to normalize not only the training set but also the validation and test set.

*B. ML Model development*

To map the extracted features to the user's gender, a binary classifier was developed using a dense neural network. This NN comprised of five hidden layers was created using Python 3.6 and Keras 2.3.1 [12] to accomplish the task of binary classification. The input to the neural network classifier was IMU signals from the sensor placed on the posterior side of the pelvis as shown in Figure 1. The output was probabilities for each class. The class with the highest probability was considered as the user's gender. The five hidden layers consisted of 500, 250, 50, 20, and 4 neurons respectively. The hidden layers used ReLU (rectified linear unit) [13] as the activation function and the output layer used Softmax as the activation function [14]. Binary cross entropy was used as a loss function with Adam (adaptive moment estimation) as the optimizer [15]. The number of hidden layers and neurons was determined by trial and error to avoid overfitting of the model. Neural networks trained using smaller batch sizes have been shown to generalize well [16]. Therefore, through trial and error, a small batch size of sixteen was used to train the classifier.

Macro-Averaged F-1 score and confusion matrix provided quantitative measures indicating the performance of the trained classifier models across the validation and test datasets. F-1 score is the harmonic mean of precision and recall which indicates the accuracy with which the classifier identifies a class and is robust to the class imbalance in dataset [18]. Precision evaluates the fraction of correctly classified instances among the ones classified as positive. The recall is a metric that quantifies the number of correct positive predictions made out of all positive predictions that could have been made. A detailed description of these assessment methods can be found in [19].

*C. Relevance identification*

Layerwise Relevance propogation was applied to the trained model and the test data set to identify importance of each frame within the 1 second time window for each axis of gyroscope and accelerometer axis [17]. Furthermore, to identify the influence of each individual axis of gyroscope and accelerometer on the gender, importance of each time frame within 1 second window was averaged to represent importance of relevant axes. Lastly, to identify the differences in the features specifically affecting the gait of males and females, the test set was divided into two parts. One part consisted of signals collected from male participants whereas the other part was comprised of signals collected from female participants. LRP was then applied to each of these parts separately.

## IV. RESULTS

Table 1 illustrates the confusion matrix which provides the number of input signals that were correctly and incorrectly classified for males and females. It can be seen that the trained model was not biased towards predicting a certain class. An F1 score of 0.795 was observed for the validation dataset and an F1 score of 0.76 was observed for the test set.

| Validation | Female | Male |
|---|---|---|
| Female | 191 | 43 |
| Male | 56 | 193 |

F1 Score: 0.795

| Test | Female | Male |
|---|---|---|
| Female | 319 | 102 |
| Male | 100 | 330 |

F1 Score: 0.765

*Table 1 Tables indicating the number of records correctly and incorrectly classified for each class by the trained classifier for validation and test set.*

Table 2 illustrates the results of applying LRP on the trained neural network classifier using different variants of the test dataset. It can be seen that the two variants of LRP identified the X-axis of the accelerometer to have the highest influence over gender. Similarly, Z-Axis was identified to have higher influence only in the overall and male dataset. Lastly, the Gradient method indicated X-Axis of Gyroscope to have a higher influence on gender.

| Dataset | Method | GX | GY | GZ | AX | AY | AZ |
|---|---|---|---|---|---|---|---|
| Overall | Gradient | 0.028984 | 0.001346 | **0.15614** | 0.014353 | 0.032904 | -0.06998 |
| | LRP | 0.061684 | -0.0158 | -0.06235 | **0.209932** | -0.01644 | **0.185028** |
| | LRP Alpha 2 Beta 1 | 0.274432 | 0.146506 | 0.091445 | **0.550122** | 0.082014 | **0.440892** |
| Male | Gradient | 0.023042 | -0.00518 | **0.076473** | 0.013318 | 0.016372 | -0.02136 |
| | LRP | 0.038603 | -0.01647 | -0.06299 | **0.087645** | -0.01751 | **0.087022** |
| | LRP Alpha 2 Beta 1 | **0.184743** | 0.070993 | 0.049662 | **0.175398** | 0.021911 | **0.192777** |
| Female | Gradient | 0.035053 | 0.008017 | **0.23751** | 0.01541 | 0.049789 | -0.11963 |
| | LRP | 0.085259 | -0.01512 | -0.0617 | **0.334834** | -0.01534 | **0.285128** |
| | LRP Alpha 2 Beta 1 | 0.366037 | 0.223634 | 0.134122 | **0.932858** | 0.143402 | **0.694311** |

*Table 2 illustrates the influence of each axis of the inertial sensor on gender. Bold values highlighted in grey indicate features with a positive influence on gender.*

## V. DISCUSSION

The primary objective of this study was to propose a framework for developing gait-based user demography estimators. A Dense Neural Network classifier was trained to identify the gender of an individual using a dataset consisting of inertial measurements collected from individuals during level-walk. Offline testing was conducted to assess the prediction accuracy of the trained classifier. Lastly, using the Layerwise Relevance Propagation (LRP) technique features contributing the most towards the classification of gender were identified. The results of our study indicate that the framework can be used to develop user demography estimators relying on the gait of an individual. We also illustrated that the limiting black-box nature of neural networks can be averted with the help of LRP. Furthermore, the proposed framework has the advantage of not requiring feature extraction or joint angle computations.

As shown in Table 1, the F1 scores of 0.76 on unseen data and confusion matrix indicate that the trained classifier had good classification accuracy and did not have any bias towards the prediction of a particular class. The LRP analysis revealed the highest influence of X-Axis of Accelerometer on predicting gender. Based on the sensor orientation and location, acceleration along the X-axis indicates the oblique motion of the pelvis. Such oblique motion in the pelvis has been identified to be a unique characteristic of female gait by numerous studies [20][21]. Furthermore, this observation coincides with greater ROM of the pelvis in the frontal plane of females [22]. This may also indicate the negative correlation of the Z-axis of the gyroscope on the gender of the individual as identified by the vanilla LRP variant, shown in Table 2. Furthermore, various studies have established slower walking speeds in females compared to male gait [23] [20]. The higher influence of Z-Axis of accelerometer identified by LRP is harmonious with the observations of walking speed made by these studies. These observations re-iterates the efficacy of the proposed framework in not only developing gait-based user demography estimators but also in troubleshooting the developed models.

Although the proposed framework yield good performance and the identified important features were harmonious to the observations of biomechanical studies, this study was limited to off-line evaluation. Therefore, future studies will be focused on empirical analysis in a real-time. Furthermore, the current study lacks comparison of LRP with various other explainable methods such as LIME [24] and SHAPE [25]. Furure studies will involve comparison of the results from these methods with LRP techniques to identify the limitations of the proposed framework. Lastly, inertial measurements were only limited to center sensor during level walking. Future studies will be focused on not only augmenting the data to include sensors on different lower body segments but also involve the effect of different walking terrains on gender classification.

## VI. Conclusion

In this paper, a novel framework for designing gait-based user demography estimatior was presented. The framework uses inertial measurements from pelvis and a neural network classifier to identify gender based on gait of an individual. Furthermore, the use of LRP to identify the important feature allows for validation and trouble shooting of the trained classifiers. The model was trained using inertial data without any post-processing such as converting IMU data in global coordinates to compute joint angles or extracting features. In the present study, the models' performance on the dataset recorded during level walk showed good performance. Classification accuracy of 76% was achieved by the neural network classifier. The observed results are quite promising, propounding the use of the proposed framework in designing classifiers to identify gender. The proposed framework can be easily extended to estimate other user demraphics such as age, body-type etc. Future work will involve real-time testing of the proposed framework with additional inertial sensors and with actual participants.

## Acknowledgment

We would like to thank OU-ISIR team for making the gait dataset easily accessible.